\documentclass[prb, aps, amsmath, amssymb, amsfonts,superscriptaddress, twocolumn]{revtex4}
\usepackage[german,american,english]{babel}
\usepackage{graphicx}
\usepackage{graphics}
\usepackage{dcolumn}
\usepackage{bm}
\usepackage{amssymb}
\usepackage{amsmath}
\usepackage{amsfonts}
\usepackage{epsfig}

\newcommand {\td} [2] {\frac{d #1}{d #2}}

 \newcommand {\beq}{\begin{equation}}
\newcommand {\eeq}{\end{equation}}
\newcommand {\beqn}{\begin{eqnarray}}
\newcommand {\eeqn}{\end{eqnarray}}
\newcommand {\bit}{\begin{itemize}}
\newcommand {\eit}{\end{itemize}}

\newcommand{\ba}{\begin{array}{rl}}
\newcommand{\ea}{\end{array}}
\newcommand{\bc}{\begin{cases}}
\newcommand{\ec}{\end{cases}}

\newcommand{\dps}{\displaystyle}

\newcommand{\om}{\iffalse}

\begin{document}
\title{Screening, Friedel oscillations and low-temperature conductivity in topological insulator thin films}
\author{Weizhe Edward Liu}
\affiliation{School of Physics, The University of New South Wales, Sydney 2052, Australia}
\author{Hong Liu}
\affiliation{ICQD, Hefei National Laboratory for Physical Sciences at the Microscale, University of Science and Technology of China, Hefei 230026, Anhui, China}
\author{Dimitrie Culcer}
\affiliation{School of Physics, The University of New South Wales, Sydney 2052, Australia}
\begin{abstract}
In thin topological insulator films, the top and bottom surfaces are coupled by tunneling, which restores backscattering and strongly affects screening. We calculate the dielectric function in the random phase approximation obtaining a closed-form result. Unlike independent TI surfaces, the dielectric function of thin films exhibits a valley as a function of wavenumber $q$ and tunneling, as well as a cusp at $q=2k_F$, with $k_F$ the Fermi wave vector. As a result of the cusp, Friedel oscillations decay with distance $r$ as $\sin(2k_Fr)/(2k_Fr)^2$. We determine the longitudinal conductivity $\sigma$ in the first Born approximation at low temperatures where screened impurities provide the dominant scattering mechanism. At high electron densities $n_e$, $\sigma \propto n_e$, while at low densities $\sigma \propto n_e^{3/2}$.
\end{abstract}
\date{\today}
\maketitle

\section{Introduction}

Considerable research has been stimulated by the unique physics of topological insulators (TIs)  \cite{Hasan_TI_RMP10}, which have conducting states on their surfaces (3D) or edges (2D) when an insulating gap exists in the bulk. Strong spin-orbit coupling enables 3D TI surfaces to support 2D spin-polarized massless Dirac fermions with a linear dispersion, exhibiting exotic phenomena with no counterpart in conventional conductors, such as a quantized anomalous Hall effect \cite{Yu_TI_QuantAHE_Science10, Chang_QAHE_exper_Science2013}. TI surfaces proximity-coupled to a superconductor may support Majorana fermions \cite{Fu_Proximity_Majorana_PRL08} enabling topological quantum computing \cite{SDS_TQC_RMP08}. Coupled spin-charge transport in TI is researched with a view to spintronics applications \cite{Schwab_TI_LokTnl_EPL11, Culcer_TI_PhysE12}, while TIs are also seen as a platform for studying the interplay of interactions and spin-orbit coupling \cite{Culcer_TI_Int_PRB11, PalYudson_TI_ee_PRB2012, Araujo_TI_PRB2013, AshrafiRashba_TI_ee_PRB2013,WangCulcer_TI_Kondo_PRB2013, NogueiraEremin_TI_X2013}.

At present, TI surface transport is obscured by bulk doping. One pathway towards reducing bulk transport has been to make TIs thinner. In a TI thin film (TITF), a layer pseudospin indexes the top and bottom surfaces. In films thinner than 6 quintuple layers \cite{Linder_TI_size_PRB2009, Liu_TIF_PRB2010, LuShan_TITF_MassiveDirac_spinphys_PRB2010, Park_TITF_topoloprotect_PRL2010, Chang_TIF_ferromagnetism_AHE_NP2013, Hirahara_TITF_transition_PRB2010, Sakamoto_TITF_transition_PRB2010, Taskin_TITF_transport_PRL2012} tunneling is enabled between the surfaces, referred to as interlayer tunneling. This gives a mass in the Dirac dispersion and, with time reversal symmetry preserved, the conduction and valence bands are each twofold degenerate. Layer pseudospin dynamics can be tuned by modulating the tunneling parameter, for example by altering the film thickness, by strain and external electromagnetic fields. 

Nontrivial surface states make TITF physics different from metal and semiconductor films. Furthermore, unlike bilayer graphene, TITFs are gapped, scattering through $\pi/2$ is allowed, and the Hamiltonian describes the real spin in a single-valley system. With the study of TITF taking off \cite{Tse_TI_MOKE_PRL2010, Tse_TIF_MOMEE_PRB2010, Zyuzin_TIF_QPT_PRB2011, Garate_TIF_WL_WAL_PRB2012, Zhang_TITF_WL_PRB2013}, understanding the interplay of spin-orbit interaction, layer pseudospin and electron-electron interactions in these structures is critical for their future functionality. Exciting predictions exist, such as topological exciton condensation \cite{Seradjeh_TIF_excitoncondens_chargefract_PRL2009, Seradjeh_TITF_TEC_PRB2012, Efimkin_TIF_EHP_PRB2012, KimHankiewicz_TI_ExcitonSuperfluid_PRB2012} and quantum Hall superfluidity \cite{Tilahun_TIF_QHS_PRL2011}. 

The dielectric function is the first step in understanding interaction physics such as Fermi-velocity renormalizations \cite{BorghiPolini_MLGBLG_Fermi_enhancement_SSC09}, Fermi liquid properties, plasmons, the Kohn anomaly \cite{HwangDasSarma_BLG_screening_PRL2008}, as well as spontaneous symmetry breaking such as ferromagnetism, which would enable seamless integration of spin and semiconducting technology. It also determines the screened impurity potentials responsible for electrical resistance at low temperatures. In this work, we calculate the TITF dielectric function in the random phase approximation (RPA), obtaining an analytic result. Tunneling has a drastic effect on screening, which takes on a distinct form from that in any other known conductor. Firstly, the screening function has a valley as a function of wave vector and tunneling. Secondly, since tunneling restores backscattering in TITF, a cusp appears at $2k_F$ in the polarizability. The cusp alters also the form of Friedel oscillations \cite{SimionGuiliani_FL_FriedelOsci_PRB2005}, which can be studied using scanning tunneling microscopy \cite{LiuQiZhang_TI_Friedel_PRB2012}. Finally, interaction effects in TITF transport remain uncharted territory \cite{Ghaemi_TIF_transport_thermoelectric_PRL2010, LuZhao_TITF_transport_PRB2013}, and we study the low-temperature TITF conductivity in the experimentally relevant metallic regime, determining its general dependence on the electron density.

The outline of this paper is as follows. In Sec.~\ref{sec:Ham} we introduce the TITF Hamiltonian and its properties. In Sec.~\ref{Sec:Limit}, we give justification of our model and show its applicability. In Sec.~\ref{Sec:screen}, we calculate the polarizability and screening function in the RPA. In Sec.~\ref{Sec:Friedel}, we obtain the form of Friedel oscillations in TITF and compare the result with monolayer and bilayer graphene. In Sec.~\ref{Sec:cond}, we calculate the longitudinal conductivity in the first Born approximation using linear response theory. Finally, in Sec.~\ref{Sec:Summary}, we summarize our results.



\section{Hamiltonian}
\label{sec:Ham}

The effective Hamiltonian at wavevector $\bm k$ is \cite{Zyuzin_TIF_QPT_PRB2011, Shen_3DTI_EffMod_NJP10}
\begin{equation}
    H_{\bm k}=A\tau_z\otimes[{\bm\sigma}\cdot({\bm k}\times\hat{\bm z})]+t\tau_x\otimes \openone,
\end{equation}
where the $\tau$ matrices act in layer pseudospin space, and the $\sigma$ matrices in electron spin space, with $\openone$ the identity. The first term describes two independent TI surfaces in which $\tau_z = \pm 1$ represents the top/bottom surfaces respectively, where $\hat{\bm z}$ is the unit vector in the direction of $ \bm z$, and the second term represents interlayer tunneling. Electronic states are described by the eigenenergy equation $H_{\bm k}{\bm\Psi}_{\bm k}=\epsilon_{\bm k}{\bm\Psi}_{\bm k}$, where ${\bm\Psi}_{\bm k}$ is the eigenstate and $\epsilon_{\bm k}$ the eigenenergy. After the direct diagonalization, the eigenenergies are $\epsilon_{{\bm k}\lambda}=\lambda\sqrt{t^2+A^2k^2}$, where $\lambda=\pm1$ corresponds to the conduction and valence band, respectively. From the TITF spectrum, the tunneling between two surfaces opens a gap $\Delta=2t$ between the conductance and valence bands, and in realistic TITFs $\Delta\sim\mathrm{60 meV}$ or $t\sim\mathrm{30 meV}$ if the thickness $d \approx 3$ quintuple layers, but $t$ can exceed 70 meV if $d$ is below 2 quintuple layers \cite{LuShan_TITF_MassiveDirac_spinphys_PRB2010}, while $A = 4\mathrm{eV\cdot\AA}$ \cite{Chang_TIF_ferromagnetism_AHE_NP2013}. We retain the $4 \times 4$ matrix description, which offers physical transparency and the future flexibility of adding random tunneling terms due to scattering.

Notably, the degeneracy $g=2$ in the TITF reflects time-reversal invariance \cite{LuShan_TITF_MassiveDirac_spinphys_PRB2010}. Under time reversal both $\bm\sigma$ and $\bm k$ reverse their signs, with the whole TITF Hamiltonian invariant. The two degenerate normalized eigenstates of $H_{\bm k}$ are
\begin{equation}\label{eigen}
    \ba
    \dps{\bm\Psi}_{{\bm k}\lambda}^{(1)}=&\dps\frac{1}{\sqrt{2}}
    \left(\lambda,-i\cos\alpha_ke^{i\theta_{\bm k}},\sin\alpha_k,0\right)^\mathrm{T},\\[3ex]
    \dps{\bm\Psi}_{{\bm k}\lambda}^{(2)}=&\dps\frac{1}{\sqrt{2}}
    \left(0,\sin\alpha_k,-i\cos\alpha_ke^{-i\theta_{\bm k}},\lambda\right)^\mathrm{T},
    \ea
\end{equation}
in which we define $\sin\alpha_k=t/\sqrt{t^2+A^2k^2}$, $\cos\alpha_k=Ak/\sqrt{t^2+A^2k^2}$, $\theta_{\bm k}=\tan^{-1}(k_y/k_x)$, and $T$ the transpose. Moreover, since any symmetry operation can transform one degenerate eigenstate into other degenerate eigenstates, ${\bm\Psi}_{{\bm k}\lambda}^{(1)}$ and ${\bm\Psi}_{{\bm k}\lambda}^{(2)}$ can be exchanged under the time-reversal operator $\Theta=C\tau_x\otimes(-i\sigma_y)$, where $C$ is the conjugate operator.

The TITF overlap factor $|{\bm\Psi}_{{\bm k}\lambda}^\dagger\cdot{\bm\Psi}_{{\bm k'}\lambda'}|^2$ is
\begin{equation}\label{TITFoverlap}
    \ba
    &\dps\left|{\bm\Psi}_{{\bm k}\lambda}^\dagger\cdot{\bm\Psi}_{{\bm k'}\lambda'}\right|^2
    =\frac{1}{2}\sum_{d_1d_2}\left|{\bm\Psi}_{{\bm k}\lambda}^{(d_1),\dagger}
    \cdot{\bm\Psi}_{{\bm k'}\lambda'}^{(d_2)}\right|^2\\[3ex]
    =&\dps\frac{1+\lambda\lambda'(\sin\alpha_k\sin\alpha_{k'}+\cos\alpha_k\cos\alpha_{k'}\cos\gamma)}{2},
    \ea
\end{equation}
where $d_1$ and $d_2$ are used to distinguish the Kramer degeneracy, and $\gamma=\theta_{\bm k'}-\theta_{\bm k}$. From eq. (\ref{TITFoverlap}), we see that the backscattering ($\gamma=\pi$) in TITF can be allowed due to the interlayer tunneling effect, as discussed in Sec. \ref{Sec:cond}. Moreover, the incompletely suppressed backscattering will give a cusp at $2k_F$ on the polarizability as showed in Sec. \ref{Sec:screen}.

\section{Limits of applicability}\label{Sec:Limit}

Two parameters quantify the strength of interactions in TITF and the limits of validity of the RPA. Firstly, we define an effective background dielectric constant $\kappa$. The physics of films is determined by $d$ and the Fermi wave vector $k_F$ \cite{Tilahun_TIF_QHS_PRL2011}. We take a Bi$_2$Se$_3$ film as an example, with $\kappa_{Bi2Se3} \approx 100$, grown on a semiconductor substrate with $\kappa_{s} \approx 11$. For $k_Fd \gg 1$, the film is thick, and the two surfaces are independent \cite{AndoFowler_2D_RMP1982}. For the top surface, where one side is exposed to air, $\kappa_{top} = (\kappa_{Bi2Se3} + 1)/2 \approx 50$. For the bottom surface $\kappa_{btm} = (\kappa_{Bi2Se3} + \kappa_{s})/2 \approx 55$. Both are independent of $d$. For $k_Fd \ll 1$, the film is ultrathin and can be approximated as a pure 2D system, with $\kappa = (1 + \kappa_{s})/2 \approx 6$, also independent of $d$. However, since the TI bulk cannot be eliminated, $\kappa \approx 6$ is an ideal lower bound. In films studied here, with $d$ of the order of 3 nm, $\kappa$ has contributions from both the TI bulk and the semiconductor substrate, and for Bi$_2$Se$_3$ can range between 6 and 55. Two experiments have extracted $\kappa \approx 30$ for relatively thick films of Bi$_2$Se$_3$ (10 nm $< d<$ 20 nm) \cite{Beidenkopf_NP11, Kim_TI_e-ph_PRL12}. Hence, $\kappa$ is treated as a phenomenological parameter to be measured separately for each film.

Secondly, the interaction parameter
\begin{equation}
r_s =\frac{e^2\sqrt{\pi n_e}}{\kappa(\varepsilon_F-t)}
\end{equation}
represents the ratio of the electrons' average Coulomb potential and kinetic energies, with $n_e$ the electron density and $\varepsilon_F$ the Fermi energy. The random phase approximation is applicable for $r_s \ll 1$ \cite{SDS_Gfn_RMP11}. We define a reference density $n_0 = t^2/(2\pi A^2)$. For $t \approx 30$ meV, $n_0 \approx 10^{11}$ cm$^{-2}$ at the lower end of densities considered in transport. For $\kappa = 30$, $ r_s \ll 1$ requires $n_e \gg 0.03 n_0$, while for $\kappa = 10$ and 6 we require $n_e/n_0 \gg 0.3$ and 1.07 respectively. Therefore, RPA is an excellent approximation for TITF at densities commonly encountered in  transport. (For example, for the case of graphene grown on SiO$_2$, $r_s \approx 0.8$, as discussed in Ref.~ \onlinecite{SDS_Gfn_RMP11}.)

We wish to dwell briefly on the relationship between $\kappa$, $d$ and $t$. As noted above, the dependence of $\kappa$ on the thickness $d$ of the film cannot, in general, be written in closed form, and $\kappa$ must be treated as a phenomenological parameter. The dependence of $t$ on $d$ is also generally only obtainable numerically \cite{LuShan_TITF_MassiveDirac_spinphys_PRB2010}. In principle, one could invert $t(d)$ to obtain $d(t)$, re-express $\kappa(d)$ as $\kappa(t)$, and write the dielectric function $\epsilon (q)$, defined below, as a function of $t$ alone (or, alternatively, as a function of $d$ alone). This approach may be useful when modeling a  specific experimental sample. It is, however, counterintuitive as well as impractical for our purposes, since it requires the full dependence of $\kappa$ on $t$. In this work, we have focused on the electronic contribution to the polarization, contained in $\Pi(q)$ (also defined below), and we have presented our main results in Figs.~\ref{TITF_intrinsic} and \ref{TITF_Extrinsic} in a form which is independent of whether $\kappa$ depends on $t$ or not (see also Sec:screen). Moreover, in the two extreme limits discussed above $\kappa$ is independent of $t$, and the only $t$-dependence in $\epsilon(q)$ comes from the electronic contribution to the polarization. 

\section{Screening and Polarizability}\label{Sec:screen}

\begin{figure}
    \includegraphics[width=0.75\columnwidth]{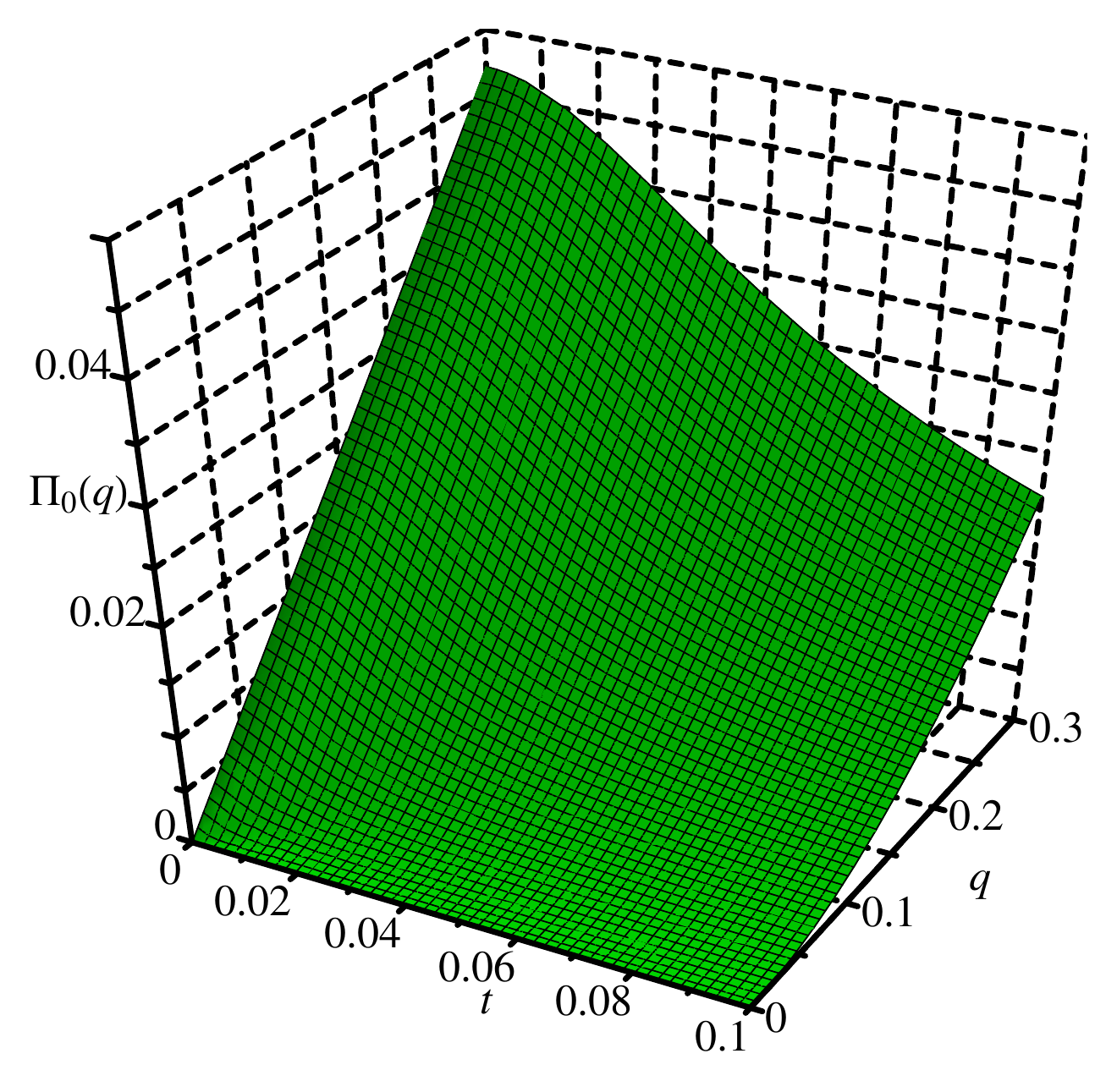}\\
    \caption{Intrinsic TITF polarizability $\Pi_0(q)$ in multiples of $\Pi_0$ (defined in the text), with $t$ and $q$ in eV and $\mathrm{nm}^{-1}$, respectively. The $t=0$ case is the TI intrinsic polarizability which is proportional to $q$, while tunneling reduces $\Pi_0(q)$ from its value for TI surfaces.}\label{TITF_intrinsic}
\end{figure}

The static TITF screening function has the form\cite{Culcer_TI_Kineq_PRB10}:
\begin{equation}
\epsilon(q)=1+\frac{2\pi e^2}{\kappa q}\Pi(q),
\end{equation}
where
\begin{equation}\label{filmpolari}
    \Pi(q)=-\frac{2}{S}\sum_{{\bm k}\lambda\lambda'}\frac{f_{{\bm k}\lambda}-f_{{\bm k'}\lambda'}}
    {\epsilon_{{\bm k}\lambda}-\epsilon_{{\bm k'}\lambda'}}
    \left|{\bm\Psi}_{{\bm k}\lambda}^\dagger\cdot{\bm\Psi}_{{\bm k'}\lambda'}\right|^2,
\end{equation}
in which ${\bm k'}={\bm k}+{\bm q}$, $f_{{\bm k}\lambda}$ is the electron distribution in band $\lambda$, and $S$ the area. Two contributions from $\Pi(q)$ are distinguished, termed intrinsic and extrinsic. To determine the intrinsic contribution, we assume $\varepsilon_F \in [-t,t]$, and, at $T=0$, $f_{{\bm k}-}=1$ and $f_{{\bm k}+}=0$. For extrinsic TITF with positive doping, when $T=0$, $f_{{\bm k}-}=1$ and $f_{{\bm k}+}=\theta(k_F-k)$, where $\theta(x)$ is the unit step function and $k_F = \sqrt{2\pi n_e}$. In general,
\begin{equation}
    \Pi_0(q)=\Pi_0\left(\frac{t}{2}+\frac{A^2q^2-4t^2}{4Aq}\tan^{-1}\frac{Aq}{2t}\right),
\end{equation}
where $\Pi_0 = 1/(\pi A^2)$. In Fig.~\ref{TITF_intrinsic}, $\Pi_0(q)$ is plotted as a function of $t$ and $q$ in units of $\Pi_0$. When $t=0$ (no tunneling) the intrinsic polarizability is linear in $q$, as for an independent TI surface. However, when $q=0$ the intrinsic polarizability vanishes because the interband overlap is zero for ${\bm k'}={\bm k}$. Remarkably, when $q\to0$ for $t\ne0$, the intrinsic polarizability is quadratic in $q$, which is drastically different from the zero tunneling case.

\begin{figure}
    \includegraphics[width=0.77\columnwidth]{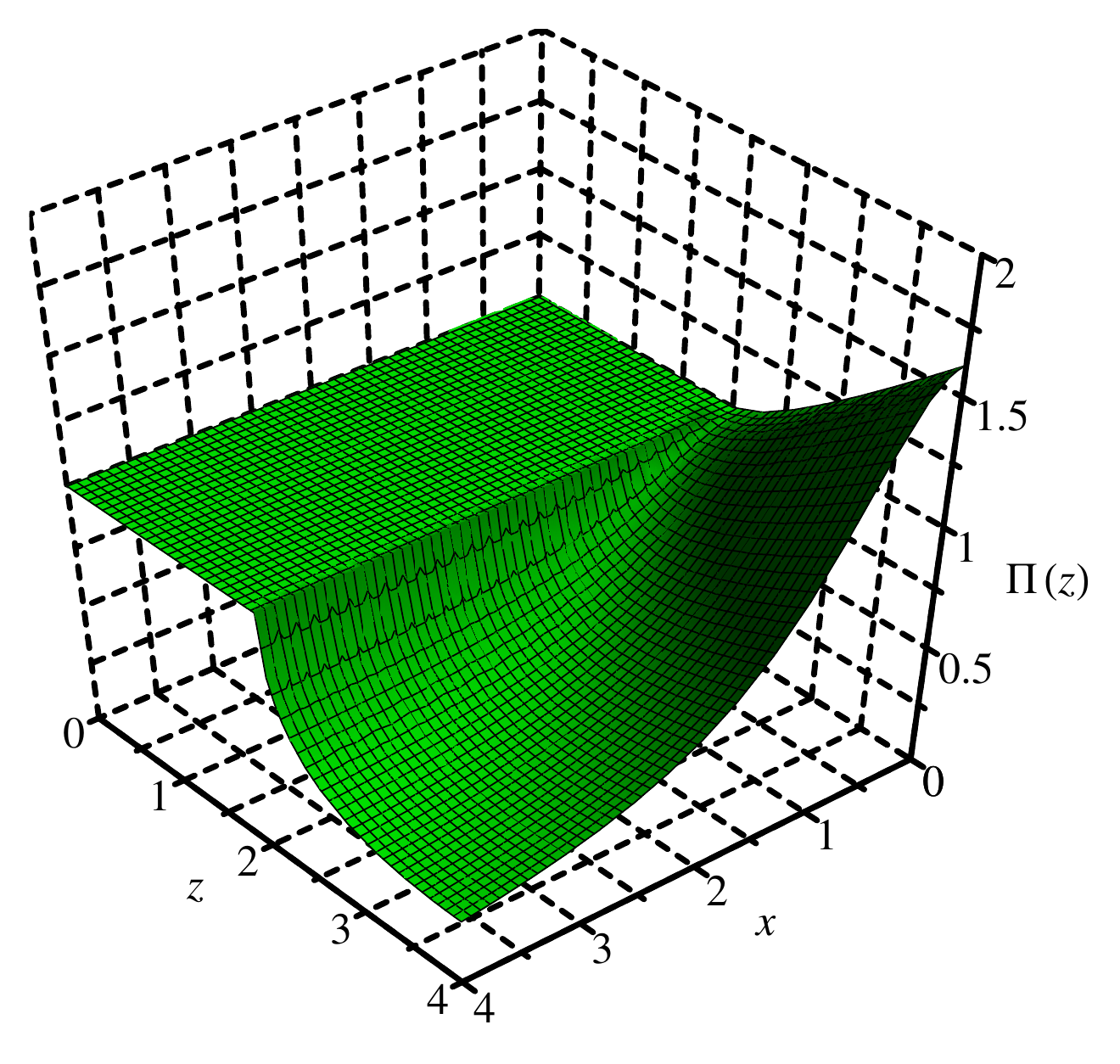}\\
    \caption{Extrinsic TITF polarizability $\Pi(z)=\Pi(q)/D_0$ as a function of $x=t/(Ak_F)$ and $z=q/k_F$. The extrinsic polarizability is fixed at 1 when $q\leqslant2k_F$, and a cusp is evident at $q=2k_F$. Note the valley in the polarizability for $q>2k_F$.}\label{TITF_Extrinsic}
\end{figure}

The extrinsic polarizability
\begin{equation}
\Pi(q)=D_0\left[1+f\left(\frac{q}{k_F}\right)\theta(q-2k_F)\right],
\end{equation}
in which $D_0 = \sqrt{t^2+A^2k_F^2}/(\pi A^2)$ is the areal density of states at the Fermi level and
\begin{equation}
f(z)=-\frac{\sqrt{z^2-4}}{2z}+\frac{z^2-4x^2}{4z\sqrt{x^2+1}}\sin^{-1}\sqrt{\frac{z^2-4}{z^2+4x^2}},
\end{equation}
where $z=q/k_F$ and $x=t/Ak_F$. In Fig.~\ref{TITF_Extrinsic}, $\Pi(z)=\Pi(q)/D_0$ is plotted as a function of $x$ and $z$. Thanks to the restoration of backscattering, a cusp exists at $q=2k_F$ for $t\ne0$, yet, at sufficiently large $q$, the extrinsic polarizability becomes approximately linear in $q$. The interplay between strong spin-orbit characteristic of individual TI surfaces and interlayer tunneling results in a valley in the TITF polarizability. For example, setting $k_F = t/A$, when $q$ lies in the neighborhood of $2k_F$, tunneling is dominant and the TITF polarizability decreases as $1-\sqrt{(q-2k_F)/4k_F}$. Yet with $q$ increasing, the spin-orbit interaction is more pronounced, and eventually $\Pi(q)$ begins to rise again.

\begin{figure}
    \includegraphics[width=0.9\columnwidth]{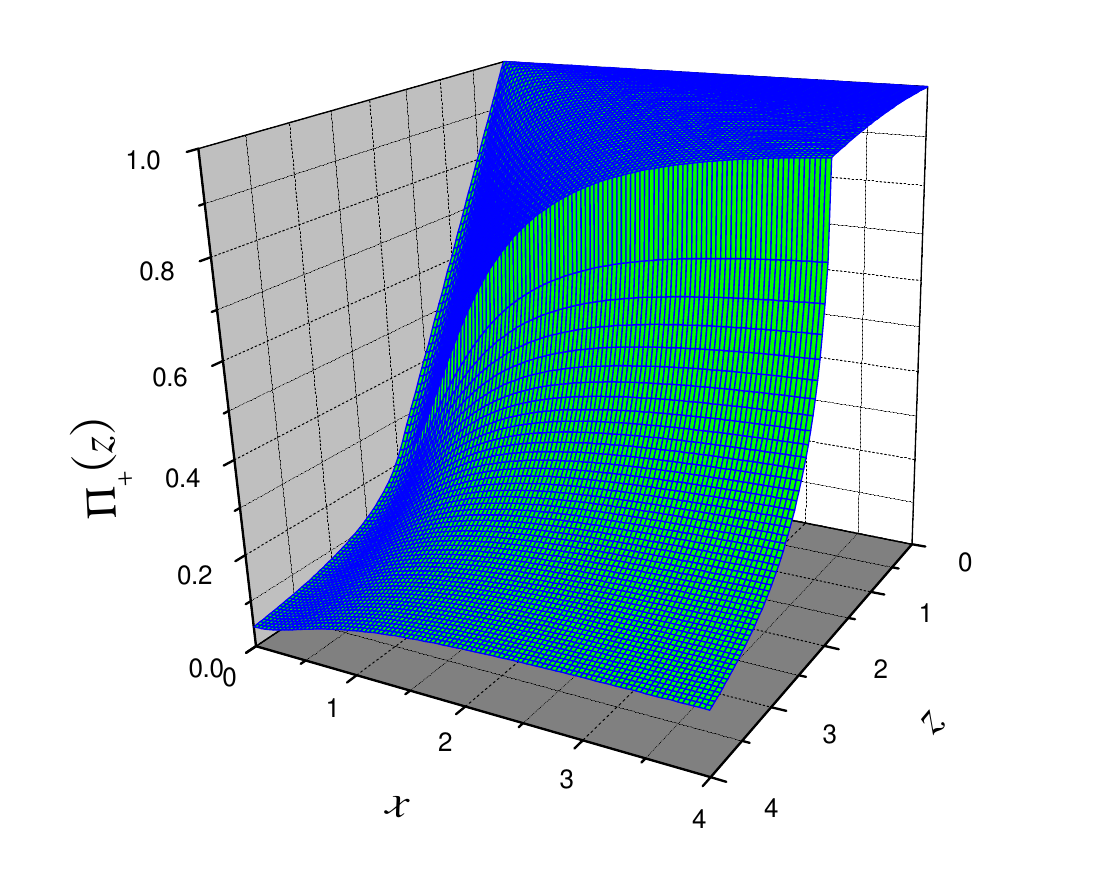}\\
    \caption{$\Pi_+(z)$ in TITF. There is also a cusp at $z=2$, and $\Pi_+(z)$ will approach zero smoothly as $q\to0$.}\label{TITF_Pi+}
\end{figure}

\begin{figure}
    \includegraphics[width=0.95\columnwidth]{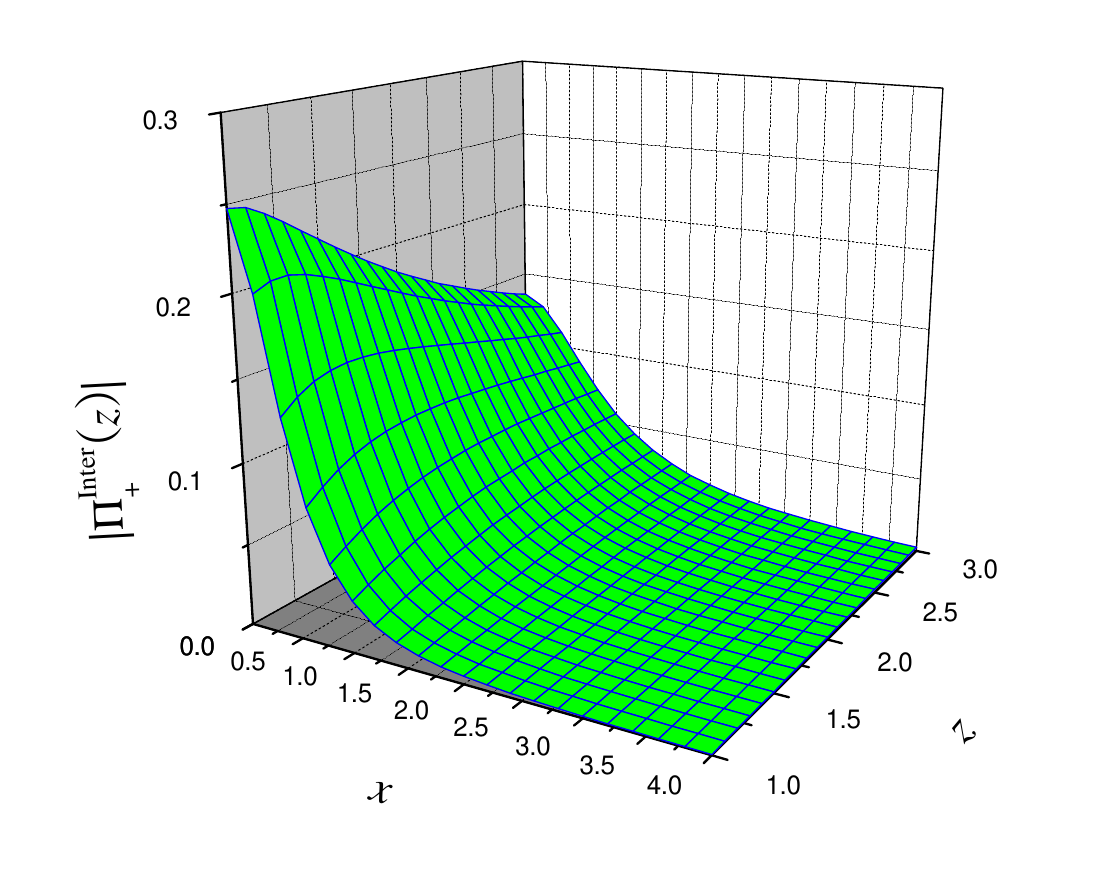}\\
    \caption{$|\Pi^\mathrm{inter}_+(z)|$ in TITF. Since there is no cusp at $z=2$ in $\Pi^\mathrm{inter}_+(z)$, the interband part of the $\Pi_+(q)$ is not related to the cusp at $z=2$.}\label{TITF_Pi+_inter}
\end{figure}

\begin{figure}
\begin{center}
    \includegraphics[width=0.75\columnwidth]{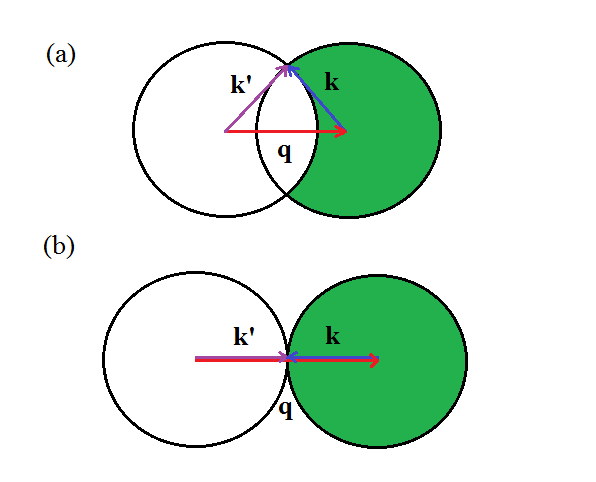}\\
    \caption{Contributing electrons (green area) for the intraband screening in (a) $q<2k_F$ case; (b) $q=2k_F$ case. When $q<2k_F$, only a part of the conduction electrons participate, and as $q$ increases, this proportion also increases. But when $q=2k_F$, the proportion saturates and there is no further increase.}\label{saturation}
\end{center}
\end{figure}

In the following, we divide the extrinsic polarizability into two parts:
\begin{equation}\label{Pi-}
    \Pi_-(q)=\frac{2}{S}\sum_{\bm k}\frac{f_{{\bm k}-}+f_{{\bm k'}-}}
    {\epsilon_k+\epsilon_{k'}}F^\mathrm{inter}_{{\bm k},{\bm k'}},
\end{equation}
\begin{equation}\label{Pi+}
    \ba
    \dps\Pi_+(q)=-\frac{2}{S}\sum_{\bm k}\bigg[&\dps
    \frac{f_{{\bm k}+}-f_{{\bm k'}+}}{\epsilon_k-\epsilon_{k'}}F^\mathrm{intra}_{{\bm k},{\bm k'}}\\[3ex]
    &\dps+\frac{f_{{\bm k}+}+f_{{\bm k'}+}}{\epsilon_k+\epsilon_{k'}}F^\mathrm{inter}_{{\bm k},{\bm k'}}\bigg],
    \ea
\end{equation}
where $\epsilon_k=\sqrt{t^2+A^2k^2}$. We define
\begin{equation}
F^\mathrm{intra}_{{\bm k},{\bm k'}}=\frac{1+(\sin\alpha_k\sin\alpha_{k'}+\cos\alpha_k\cos\alpha_{k'}\cos\gamma)}{2}
\end{equation}
and
\begin{equation}
F^\mathrm{inter}_{{\bm k},{\bm k'}}=\frac{1-(\sin\alpha_k\sin\alpha_{k'}+\cos\alpha_k\cos\alpha_{k'}\cos\gamma)}{2}
\end{equation}
as the intraband and interband overlap factors, respectively. Actually, the $f_-$ part $\Pi_-(q)$ is equivalent to the intrinsic polarizability $\Pi_0(q)$ in which the intraband overlap is extinct. $\Pi_+(q)$ is the $f_+$ part of the extrinsic polarizability,
\begin{widetext}
\begin{equation}
    \Pi_+(z)=\frac{\Pi_+(q)}{D_0}=
    \bc
    \dps1-\frac{x}{2\sqrt{x^2+1}}-\frac{z^2-4x^2}{4z\sqrt{x^2+1}}\tan^{-1}\frac{z}{2x}&\dps z\leqslant2,\\[3ex]
    \dps1-\frac{\sqrt{z^2-4}}{2z}-\frac{x}{2\sqrt{x^2+1}}+\frac{z^2-4x^2}{4z\sqrt{x^2+1}}\left[
    \sin^{-1}\sqrt{\frac{z^2-4}{z^2+4x^2}}-\sin^{-1}\sqrt{\frac{z^2}{z^2+4x^2}}\right]
    &\dps z>2.
    \ec
\end{equation}
\end{widetext}
In Fig.~\ref{TITF_Pi+} we plot $\Pi_+(z)$ as a function of $x\in[0,4]$ and $z\in[0,4]$. As expected, there is also a cusp of $\Pi_+(z)$ at $z=2$. The physics underlying the $2k_F$ cusp is explained as follows. Firstly, the cusp at $z=2$ does not involve the interband part in Eq. (\ref{Pi+}) which is denoted as $\Pi^\mathrm{inter}_+(q)$ below. For the innocence of $\Pi^\mathrm{inter}_+(q)$, we plot $|\Pi^\mathrm{inter}_+(z)|=|\Pi^\mathrm{inter}_+(q)|/D_0$ for $x\in[0,4]$ and $z\in[1,3]$ in Fig.~\ref{TITF_Pi+_inter}, and in Fig.~\ref{TITF_Pi+_inter}, we see that there is no cusp at $z=2$. Thus, the intraband part of $\Pi_+(q)$ is entirely responsible for the cusp. From Eq. (\ref{Pi+}), the intraband part can be expressed as
\begin{equation}\label{Pi+intra}
    \Pi^\mathrm{intra}_+(q)=-\frac{2}{S}\sum_{\bm k}
    \frac{f_{{\bm k}+}-f_{{\bm k'}+}}{\epsilon_{k}-\epsilon_{k'}}F^\mathrm{intra}_{{\bm k},{\bm k'}}.
\end{equation}
From Eq. (\ref{Pi+intra}), the numerator $f_{{\bm k}+}-f_{{\bm k'}+}$ indicates that only the green area in the Fig.~\ref{saturation} contributes to the intraband polarizability. When $q<2k_F$, only a part of the conductance electrons participate in the intraband polarizability, and with $q$ increasing, the number of the involved electrons will also increase. But when $q=2k_F$, all conductance electrons are included, thus the effect of the factor $f_{{\bm k}+}-f_{{\bm k'}+}$ is saturated. Moreover, from eq. (\ref{TITFoverlap}) the backscattering in TITF is not completely suppressed in the intraband case if $t\ne0$, so the intraband polarizability should decline sharply when $q>2k_F$. These are the reasons why there is a cusp in $\Pi_+(q)$, as well as in $\Pi(q)$, although the cusp is not obvious when $t<0.5Ak_F$ ($x<0.5$). Moreover, we see that $\Pi_+(q)$ in TITF will approach zero in a smooth way when $q\to\infty$.

We wish to highlight the fact that, in Figs. \ref{TITF_intrinsic} and \ref{TITF_Extrinsic}, we have plotted the polarization $\Pi(q)$, which does {\it not} depend on $\kappa$. It is therefore irrelevant in these figures whether $\kappa$ depends on $d$ or not. If $\kappa$ depends on $d$, this will be reflected in additional $d$-dependence of $\epsilon(q)$. For example, in determining Friedel oscillations below, special care must be taken, as we endeavor to explain, to isolate the thickness dependence of $t$ from the the thickness dependence of $\kappa$. However, the thickness dependence of $\kappa$ will not alter the shape of Figs. \ref{TITF_intrinsic} and \ref{TITF_Extrinsic}, which are valid for any $\kappa$.  

\section{Friedel oscillations}\label{Sec:Friedel}

The static dielectric function is given by
\begin{equation}
\epsilon(q)=1 + \frac{q_\mathrm{TF}}{q}\, \bigg[1+ f\bigg(\frac{q}{k_F}\bigg)\theta(q-2k_F)\bigg], 
\end{equation}
where $q_\mathrm{TF}=2e^2\sqrt{t^2+A^2k_F^2}/\kappa A^2$ is the Thomas-Fermi wavenumber. The dielectric function has a cusp at $q=2k_F$, and a simple form for $q\leqslant2k_F$, i.e. $\epsilon(q\leqslant2k_F)=1+(q_\mathrm{TF}/q)$. The cusp affects screened Coulomb potentials in real space by giving rise to Friedel oscillations. When a point charge $Ze$ is placed on one of the two coupled topological surfaces, the leading oscillation term in its screened potential at large in-plane distances $r$ is
\begin{equation}\label{Friedel}
    \varphi(r)\sim-\frac{4Zek_F^2t^2q_\mathrm{TF}}{\kappa(A^2k_F^2+t^2)(2k_F+q_\mathrm{TF})^2}\frac{\sin(2k_Fr)}{(2k_Fr)^2}.
\end{equation} 

Independent TI surfaces correspond to the zero-tunneling limit of TITFs and, as expected, the screening and polarizability functions in TITFs reduce to the independent TI case as $t \rightarrow 0$. The intrinsic TITF polarizability is suppressed by tunneling everywhere: obviously the enhanced TITF energy denominator contributes to this suppression. Moreover, the extrinsic polarizability has a cusp at $2k_F$ due to the tunneling-induced backscattering. In contrast, for independent TI surface states, since backscattering ($q=2k_F$) is suppressed the TI polarizability has no cusp at $2k_F$. As for Friedel oscillations, when $t=0$ the term $\sin(2k_Fr)/(2k_Fr)^2$ vanishes, and the term of next highest order in $1/(k_Fr)$ yields the independent TI result  that is proportional to $\cos(2k_Fr)/(2k_Fr)^3$, which has the same qualitative form as monolayer graphene (MLG) \cite{Bena_PRL2008}. Detecting Friedel oscillations requires a regime in which $\kappa$ is independent of $d$, hence $k_Fd\ll1$ or $k_Fd\gg1$. For $n_e$ = $10^{14}$ $\mathrm{cm}^{-2}$ and $d$ = 3 $\mathrm{nm}$, $k_Fd\approx 10\gg1$ and RPA is applicable if $\kappa$ is bigger than 3 (true in all known samples). In this parameter range, one could monitor the change in the functional form of the oscillations for progressively thinner films, while measuring $\kappa$ through e.g. optical experiments. Or alternatively, by straining a film, one could change $t$ without changing $\kappa$. Thus $t$ can be extracted experimentally knowing the dependence of $\varphi(r)$ on $t$ given by Eq. (\ref{Friedel}).

Indeed, MLG shares many properties with TI surfaces since the MLG and TI Hamiltonians are connected by a unitary transformation. The TI spin is mapped onto the sublattice pseudospin of MLGs. This results in similar expressions for the screening function \cite{Ando_MLG_screening_JPSJ2006, Polini_graphene_screening_SSC2007, PoliniAsgari_MLG_screening_PRB2008, Hwang_Gfn_Screening_PRB07}, polarizability, Friedel oscillations, and conductivity \cite{Hwang_Gfn_Screening_PRB07}. Bilayer graphene (BLG) is described by a different model from TITF: interlayer tunneling couples electrons of \textit{opposite} pseudospins, whereas in TITF it couples electrons of the same spin. As a result, TITFs are gapped whereas BLG is gapless. In contrast to TITF, $\Pi_0(q)$ in BLG is independent of $q$ \cite{HwangDasSarma_BLG_screening_PRL2008}, and $\Pi(q)$ is not constant when the wavenumber is smaller than $2k_F$. However, both BLG and TITF have a cusp at $2k_F$ in the extrinsic polarizability as $d\Pi(z)/dz\propto1/\sqrt{z^2-4}\to-\infty$ in the limit $z\to2^+$ \cite{HwangDasSarma_BLG_screening_PRL2008} (similar to that of a 2D electron gas \cite{AndoFowler_2D_RMP1982}). Surprisingly therefore, Friedel oscillations in TITF have the same $r$-dependence as BLG at large $r$.\cite{HwangDasSarma_BLG_screening_PRL2008} The similarity indicates the screening function has a similar behavior in the two cases in the neighborhood of the cusp. 

\section{Conductivity and its density dependence}\label{Sec:cond}

We now discuss electron transport in the metallic regime, with $\varepsilon_F \tau/\hbar \gg 1$, where $\tau$ is the momentum relaxation time. With $\hat{H}$ the total Hamiltonian, the density operator $\hat{\rho}$ satisfies the quantum Liouville equation
\begin{equation}
\td{\hat{\rho}}{t} + \frac{i}{\hbar} \, [\hat{H}, \hat{\rho}] = 0.
\end{equation}
TITFs have double degenerate conductance and valence bands, however, in our case of positive doping, we only choose the degenerate conductance eigenstates for our bases to build a simplified two-band model in the study of transport properties, because this model is adequate for the low-temperature transport phenomenon which  only relies on the Fermi level. We project this equation onto the conduction band states $|{\bm k},{\bm\Psi}_{{\bm k},1}^{(d)}\rangle=|{\bm k}\rangle|{\bm\Psi}_{{\bm k},1}^{(d)}\rangle$, where $d$ labels the degenerate bands. We consider impurities randomly distributed throughout the film, whose potential does not couple the two layers, and are concerned with the configurational average of the impurity potential. Though the potential of a single impurity may have a layer-index dependence, it is evident by symmetry that the configurational average is independent of layer index. We are thus justified in using a single $\kappa$ to describe impurity scattering. The band Hamiltonian $\hat{H}^{dd'}_{{\bm k}{\bm k'}}$, the configurationally averaged impurity potential $|U^{dd'}_{{\bm k}{\bm k'}}|^2$, and the interaction with the external electrical field $\hat{H}^{\mathrm{E},dd'}_{{\bm k}{\bm k'}}$ are
\begin{equation}
\ba
\dps\hat{H}^{dd'}_{{\bm k}{\bm k'}}&\dps=\epsilon_k\delta_{{\bm k}{\bm k'}}\delta_{dd'},\\[3ex]
\dps |U^{dd'}_{{\bm k}{\bm k'}}|^2&\dps=\frac{n_i|U_{{\bm k}{\bm k'}}|^2}{V}F^\mathrm{intra}_{{\bm k},{\bm k'}}\delta_{{\bm k}{\bm k'}}\delta_{dd'},\\[3ex]
\dps\hat{H}^{\mathrm{E},dd'}_{{\bm k}{\bm k'}}&\dps=ie{\bm E}\cdot\left(\frac{D}{D{\bm k}}\right)_{dd'}\delta_{{\bm k}{\bm k'}},
\ea
\end{equation}
where $|U_{{\bm k}{\bm k'}}|^2=\{2\pi Ze/[\kappa_\mathrm{TI}\epsilon(|{\bm k}-{\bm k'}|)|{\bm k}-{\bm k'}|]\}^2$, $D/D{\bm k}$ is the covariant derivative of $\bm k$ that will eventually reduce to $(\partial/\partial{\bm k})\openone$, since we only consider the leading contribution to the conductivity, $n_i$ and $Z$ is the density and charge number of the impurities, respectively. We assume the temperature $T=0$ for simplicity, thus electron-phonon and electron-electron scattering can be neglected. The dominant scattering mechanism is expected to be due to ionized impurities, since typical surface roughness fluctuations average $1-2\, \AA$ whereas the centroid of the wave function is $\approx1$ nm inside the material, and is more sensitive to ionized impurities residing inside the material than to roughness.

\begin{figure}[tbp]
    \includegraphics[width=0.8\columnwidth]{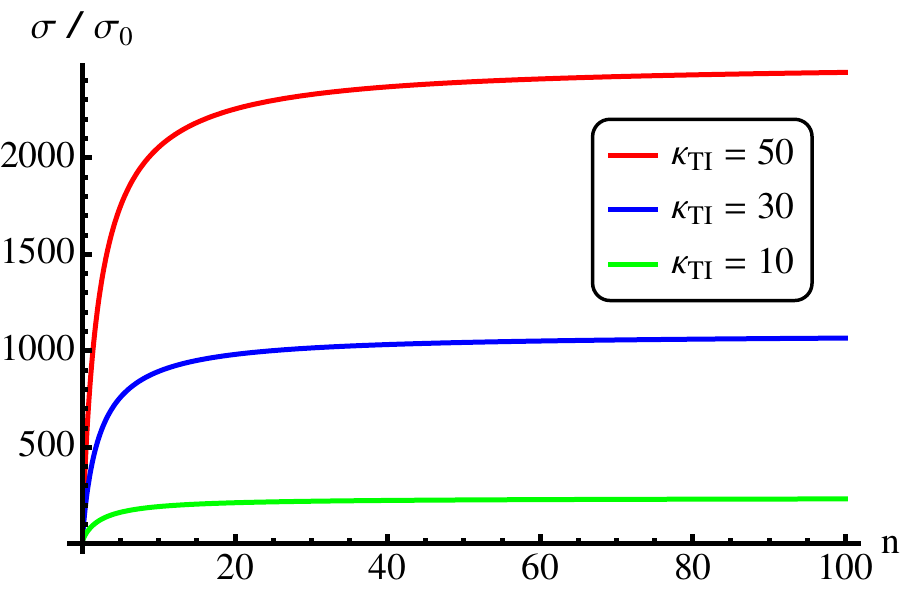}\\
    \caption{Low-temperature TITF conductivity $\sigma$, in the unit of $\sigma_0$ (having linear density dependence), as a function of the relative electron density $n$. At large $n\gg1$, all curves tend to constants, yet when $n\ll1$, they are proportional to $ n^{1/2}$.}\label{TITF_transport}
\end{figure}

We focus on linear electrical response in the first Born approximation, and the kinetic equation is\cite{Culcer_Gfn_Transp_PRB08}
\begin{equation}\label{TITF_trans}
    \frac{df_{\bm k}}{dt}+\frac{i}{\hbar}\left[H_{\bm k},f_{\bm k}\right]+\hat{J}(f_{\bm k})
    =\frac{e{\bm E}}{\hbar}\cdot\left[\frac{D}{D{\bm k}},f_{\bm k}\right],
\end{equation}
where the scattering term is
\begin{equation}\label{TITF_scattering}
    \ba
    \dps\hat{J}(f_{\bm k})=&\dps\frac{n_i\epsilon_k}{4\pi A^2\hbar}\int_0^{2\pi}|U_{{\bm k}{\bm k'}}|^2(f_{\bm k}-f_{\bm k'})d\theta_{\bm k'}\\[3ex]
    &\dps\times(1+\sin^2\alpha_k+\cos^2\alpha_k\cos\gamma).
    \ea
\end{equation}
The scattering term shows that backscattering is allowed, although only in small proportions. From eq. (\ref{TITF_scattering}), the amount of backscattering is proportional to $(t/Ak_F)^2$, and for a realistic sample, $t=30\mathrm{meV}$ and $k_F=10^8\mathrm{cm}^{-1}$, we have $(t/Ak_F)^2\approx6\times10^{-5}$. In the linear response of the electrical field, $f_{\bm k}$ in the right side of eq. (\ref{TITF_trans}) becomes $f_{0{\bm k}}\openone$, where $f_{0{\bm k}}$ is the Fermi-Dirac distribution $f_0(\epsilon_{\bm k})$ at zero temperature. We simplify $f_{\bm k}$ into a diagonal matrix $f^d_{\bm k}\openone$ in the scattering term, which is the dominant part of the density matrix deviation due to the applied electrical field, and since we are considering the steady case, the time-derivative term in eq. (\ref{TITF_trans}) is zero. The leading term of the longitudinal conductivity
\begin{equation}
\sigma=e^2Ak_F\tau/(8\pi\hbar^2)=\sigma_0D(n),
\end{equation}
where $\tau=2\hbar Ak_FD(n)/(Z^2e^4\pi n_i)$ is the momentum relaxation time, $\sigma_0=A^2n_e/(2\pi\hbar n_iZ^2e^2)$, the relative electron density $n=n_e/n_0$, and
\begin{widetext}
\begin{equation}
D(n)=\kappa^2\Bigg\{\frac{\pi}{2}\left(\alpha-6\alpha_\mathrm{TF}^2\right)+6\alpha_\mathrm{TF}-\frac{\alpha_\mathrm{TF}}{\alpha_\mathrm{TF}^2-1}(\alpha-1)+\frac{\alpha_\mathrm{TF}\sec^{-1}\alpha_\mathrm{TF}}{(\alpha_\mathrm{TF}^2-1)^{3/2}}\left[6\alpha_\mathrm{TF}^4-(9+\alpha)\alpha_\mathrm{TF}^2+2\alpha+2\right]\Bigg\}^{-1},
\end{equation}
\end{widetext}
in which
\begin{equation}
\alpha_\mathrm{TF}=\frac{q_\mathrm{TF}}{2k_F}=\frac{e^2}{\kappa A}\sqrt{1+\frac{1}{n}},
\end{equation}
and $\alpha=1+2/n$. Mathematically, when $\alpha_\mathrm{TF} < 1$,
\begin{equation}
\frac{\sec^{-1}\alpha_\mathrm{TF}}{(\alpha_\mathrm{TF}^2-1)^{3/2}} \rightarrow -\frac{\cosh^{-1}(1/\alpha_\mathrm{TF})}{(1-\alpha_\mathrm{TF}^2)^{3/2}},
\end{equation}
hence the result is valid for all $\alpha_\mathrm{TF}>0$. In Fig.~\ref{TITF_transport}, $\sigma/\sigma_0$ is shown as a function of $n$ for three trial values $\kappa$ = 10, 30, 50. When $n_e \gg n_0$ (i.e. $n\gg1$), the conductivity $\sigma$ is linear in $n_e$, though when $n_e \ll n_0$ (i.e. $n\ll 1$), $D(n)\sim\kappa^2\sqrt{n}/[6a-(2/a)]$ where $a = e^2/(\kappa A)$, so the conductivity is proportional to $n_e^{3/2}$. The density dependence of $\sigma$ for $n_e\gg n_0$ is the same as independent TI surfaces \cite{Culcer_TI_Kineq_PRB10}, since at large $k_F$ spin-orbit interaction dominates over tunneling. We expect a similar non-equilibrium renormalization of $\sigma$ as in BLG \cite{Liu_BLG_ee_PRB2013}. At low densities, of the order of $n_0$, the behavior of TITF diverges drastically from that of independent TI surfaces. Caution must be exercised, however, when extrapolating transport results to the low-density regime, since at low enough densities the sample enters the disordered regime, where transport is diffusive \cite{Kronig_TITF_PRB2013}. The results are valid as long as $\varepsilon_F\tau/\hbar \gg 1$.

\section{Summary}\label{Sec:Summary}

In summary, we have calculated the dielectric function of TITF in the RPA and shown that, unlike TI surfaces, it has a cusp at $2k_F$. The intrinsic polarizability is suppressed by the interlayer tunneling, while the extrinsic polarizability has a valley at $q > 2k_F$. Friedel oscillations take the form $\sin(2k_Fr)/(2k_Fr)^2$, compared with $\cos(2k_Fr)/(2k_Fr)^3$ in independent TI surfaces, and, in the first Born approximation, the longitudinal conductivity is proportional to $n_e$ at high densities ($n_e\gg n_0$), while $n_e^{3/2}$ at low densities ($n_e\ll n_0$). The screening function also governs the form of the RKKY interaction, where similar behavior is expected. These findings will help experiment to identify signatures of surface transport, and pave the way for the study of interaction-induced instabilities. 

\section{Acknowledgments}

We are grateful to Euyheon Hwang, Uli Zuelicke, Sankar Das Sarma, Roland Winkler, Ewelina Hankiewicz, Allan MacDonald, Haizhou Lu, Anton Burkov, and Xiaolin Wang for enlightening discussions.

\end{document}